\documentclass[12pt]{article}
\usepackage{graphicx} 
\usepackage[noadjust]{cite}

\newcommand{\beq}{\begin{equation}}
\newcommand{\eeq}{\end{equation}}
\newcommand{\bea}{\begin{array}}
\newcommand{\eea}{\end{array}}
\newcommand{\bey}{\begin{eqnarray}}
\newcommand{\eey}{\end{eqnarray}}

\newcommand{\p}{\partial}
\newcommand{\gr}{\includegraphics}
\usepackage{color}

\begin{document}

\begin{titlepage}
\vskip 2cm
\begin{center}
{\bf \Large {{The Superposition of Two Identical States: The Empty State} }}

 \vskip 10pt
{\bf
Anas Othman$^{a,b,1}$    \\}
\vskip 5pt

 \vskip 2pt
  {\sl $^a$  Department of Physics,
University of Waterloo 200 University Avenue West, Waterloo, ON, N2L 3G1, Canada }\\
{\sl $^b$Department of Physics, Faculty of Science, Taibah University\\
 Al Madinah Al Munawwarah, Saudi Arabia\\}

{\sl $^1$ a2othman@uwaterloo.ca, aothman@taibahu.edu.sa}

\end{center}

\vskip 2pt

\vskip 2pt

\begin{abstract}
In this paper we define and study a new class of states (the empty states). These states are the superposition of two identical states (self-superposition state). We defined three different representations of theses states, namely,  the source, the operator, and the self-superposition representations. Then,  we apply the empty state to an elementary example state and find its three different representations.  We apply the empty state to Fock states and find an   undetermined  state.  Then, we apply  the empty state to the coherent state to produce the empty state of the coherent state (the EC state),  and determine some of its mathematical, statistical and nonclassical properties.
\end{abstract}

\bigskip

{\em Keywords: superposition states \, \, coherent states \, \,  nonclassical states \, \, new states \, \,  }  

{\em PACS:}

\end{titlepage}

\newpage

\setcounter{footnote}{0} \setcounter{page}{1} \setcounter{section}{0} %
\setcounter{subsection}{0} \setcounter{subsubsection}{0}

%%%%%%%%%%%%%%%%%%%%%%%%%%%%%%%%%%%%%%%%%%%%%%%%%%%%%%%%%%%%%%%%%%%%%%%%%%%%%%%%%%%%%%%%%%%%%

\section{ Introduction} 
Many states have been introduced in quantum optics to observe some nonclassical effects \cite{11,75}, that  are generally important for many applications in quantum optics, (for example see \cite{9,10}). A nonclassical state can be determined in many ways, such as finding quantum interferences in the photon number distribution or finding some negative regions in the Wigner function, etc \cite{75,u1,u2}.  An example of a nonclassical state is the Fock state or number state \cite{u3}. 

 One class of nonclassical states is the superposition of two states. One possible formula for these states is $|\psi \rangle= C ( |r \rangle+ e^{i \phi} | r e^{i \theta} \rangle ) $, where the state $|r \rangle$ can be any state with a continuous variable $r$, phase parameters $\phi$ and  $\theta$, and $C$ is the normalization constant.  If  state $| r \rangle$ is the Glauber coherent state \cite{Glauber} and $\theta$ equals $\pi$, then the  state $|\psi \rangle $ becomes the Schr\"ordinger cat states which perform some nonclassical properties  \cite{Roy, eca, Yurk, Gea, Brun, Geryy}. Another example is where $| r \rangle$ can be the squeezed coherent state, where here $r$ refers to two  independent parameters, we will have the superposition of two squeezed coherent states  \cite{Yuri, oth, squ}. 

Aside from  state $| \psi \rangle$, there are more general formulas to use to construct the superposition of two states (for example see \cite{squ,  sup1, sup2, sup3, sup4, sup5, sup6, sup7, sup8, sup9}). Although these general formulas are studied extensively, there are special cases in which no physics or at least no quantum optics has yet been attributed to them, to the best of our knowledge.  These special cases are the superposition of two identical states or {\em self-superposition states}.  When $\theta$ equals zero in the above $| \psi \rangle$, we can get a possible construction of  a self-superposition state
\beq \label{ii1}
 |\psi \rangle= C( |r\rangle+e^{i \phi}|r \rangle).
 \eeq  
Another possible construction is 
\beq  \label{0z}
|\psi \rangle= \lim_{\theta \to 0} C(|r \rangle +e^{i \phi} |r e^{i \theta} \rangle), 
\eeq
and there are many other possibilities as well. 

In general, these different constructions of a self-superposition state are not equivalent. On other words, their resultant states are not same. The possible outcomes of these different constructions  can thus be classified as two different types. The first type gives the state $| r \rangle$ multiplied by a coefficient. The second type gives a new state which cannot be described in terms of the state $| r \rangle$.  The first type is trivial since we start with a state and end up with the same state. Also, because we know that the superposition of two states produces a new state, the self-superposition state should also give a new state. Therefore, we cannot consider the first type to be a self-superposition state. The first type looks like  a normal addition, and all observables will be invariant because we just altered the amplitude.

 Now, the second type which gives a new state is truly what we can call the self-superposition. To further distinguish  between the first and second types, we call the second type of state an {\em empty state}.  The detailed discussion regarding self-superposition states and the reason why they are named empty states, is discussed in the next section.

   In the following sections, we introduce the general formulation of the empty states and define three different representations to describe them: the source , the operator and the self-superposition representations. Then, we apply  the empty state to an example state and to the Fock states (empty-Fock states). We also apply the empty state to the coherent state (the EC state) and study its mathematical, statistical and nonclassical properties.

%%%%%%%%%%%%%%%%%%%%%%%%%%%%%%%%%%%%%%%%%%%%%%%%%%%%%%%%%%%%%%%%%%%%%%%%%%%%%%%%%%%%%%%%%%%%%%%%%%%%%%%%%%%%%%%%%%%%%%%%%%%%%%%%%%%%%%%%%%%%%%%%%%%%%%%%%%%%%%%%%%%%%%%%%%%%%%%%%%%%%%%%%%

\section{The Empty States} 
This section examines the  different constructions of  the self-superposition states, and the properties of empty states. Let us start by studying the possible constructions of self-superposition states, the first of which is the construction in Eq.(\ref{ii1})
\[
|\psi \rangle= C ( |r \rangle+ e^{i \phi} | r  \rangle ),
\]
where the states $| r \rangle $ and $|\psi \rangle$  are assumed to be normalized.  The normalization constant $C$ reads 
\beq  \label{i2} 
C = \frac{1}{\sqrt{2 ( 1 + \cos\phi)}}
  \eeq 
 Then, the state $| \psi \rangle $ yields 
 \beq \label{i3}
| \psi \rangle = \frac{ 1+ e^{i \phi} }{\sqrt{2(1+ \cos\phi)}} |r \rangle = e^{i \phi^\prime} | r \rangle,
\eeq
where $\phi^\prime$ is given by 
\beq
 \tan \phi^\prime= \frac{\sin\phi}{1+\cos\phi}. 
 \eeq
  The state $| \psi \rangle$ is simply the state $| r \rangle $ with a different amplitude, so this construction is trivial, and we cannot call it a self-superposition state. 
   
 Another possible construction is found  in Eq.(\ref{0z})
\[
 | \psi \rangle = \lim_{\theta \to 0} C (|r\rangle + e^{i \phi}| r e^{i \theta} \rangle ).
\]
   Neglecting the limit,  the normalization constant $C$ is given as 
\beq  \label{eee4}
C= \frac{ 1}{\sqrt{2+e^{i \phi} \langle r   | r e^{i \theta} \rangle+ e^{-i \phi} \langle r e^{i \theta} | r \rangle}}. 
\eeq
 Because we required $\theta$ to go to zero, we can expand the inner product around $\theta \approx 0$ as 
 \beq 
 \langle r | r e^{i \theta} \rangle \approx 1+ g_1(r) \theta+ g_2(r) \theta^2+\cdots,  \, \, \, \, \, \, \, \, \, \, \,   \langle r e^{i \theta} | r  \rangle \approx 1+ g_1^*(r) \theta+g_2^*(r) \theta^2+\cdots,  
 \eeq
 where $g_1(r)$ and $g_2(r)$ are functions that depend on the state $| r \rangle$. This expansion gives $1$ when $\theta$ equals $0$, thus satisfying the normalization condition of $| r \rangle$. The normalization constant can be rewritten as 
 \beq \label{jp}
 C= \frac{ 1}{\sqrt{2+2 \cos \phi + e^{i \phi} (g_1 \theta+ g_2  \theta^2)+ e^{- i \phi}(g_1^* \theta+g_2^* \theta^2 ) }}. 
 \eeq
 This normalization constant, $C$, is valid as long as $\theta \ll 1$. Now, let us expand the state $| r e^{i \theta} \rangle$, which may be expanded  as 
 \beq 
 | r e^{i \theta} \rangle \approx | r \rangle + i r \frac{ d | r \rangle }{d r} \theta +\cdots, 
 \eeq
 where we have treated the state $|r e^{i \theta} \rangle$ as a function. Then, the state $| \psi \rangle $ in Eq.(\ref{0z}) becomes 
 \beq \label{nmn}
 | \psi \rangle = \frac{|r \rangle + e^{i \phi} \left(| r \rangle + i r \frac{ d | r \rangle }{d r} \theta \right) }{  \sqrt{2+2 \cos \phi + e^{i \phi} (g_1 \theta+ g_2  \theta^2)+ e^{- i \phi}(g_1^* \theta+g_2^* \theta^2 )  }}. 
 \eeq
 Consequently, if we set $\theta$ to equal zero, we get exactly Eq.(\ref{i3}). Therefore, this state indicates that we obtained the same result as the previous construction, which thus yields to be a trivial case. Nevertheless, if and only if  the phase $\phi$ becomes equal to $\pi$, and then $\theta$ goes to zero,  we will get zero. However, if we keep $\phi=\pi$ but assuming that $g_1+g_1^*$ is zero, the state $| \psi \rangle$ becomes finite and gives 
 \beq \label{oip}
 \lim_{\theta \to 0} |\psi_{\phi =\pi} \rangle = \frac{ -i r \frac{d | r \rangle}{d r} \theta}{\sqrt{g_2+g_2^*} \,  i \theta}= -\frac{r}{\sqrt{g_2+g_2^*}} \frac{ d | r \rangle}{d r}.
 \eeq
  This result gives a new state other than $| r\rangle$, since the derivative of a state  generally gives a new state. It means that the construction in Eq.(\ref{0z}) is not trivial and constructs a self-superposition state. As discussed in the introduction, we give this type of states a special name which is the empty state. Before we progress,  we want to point out certain important remarks. Making $\theta$ go to zero is a requirement of constructing two identical states. Also, we found that phase $\phi $ should be $\pi$ for us to get a new state out of $| r \rangle$. In other words, if $\phi$ has other values rather than $\pi$, we would not be able to find a new state. Also, note that $g_1+g_1^*$ has to be zero in order to get a nonvanishing state, so this requirement is a necessary condition for constructing the empty state.

  The reason why we name these states as empty states is because in order to construct these states, the phase $\phi$ has to equal $\pi$. Now, in Eq.(\ref{0z}) when we take $\theta$ to zero, and $\phi$ equals to $\pi$, we subtract two identical states as follows: 
 \beq \label{hh}
 | \psi \rangle = \lim_{\theta \to 0} C(|r \rangle -|r e^{i \theta} \rangle). 
 \eeq
 Intuitively,  subtracting  two identical states should give zero.  Because of this first intuition, I named these states  the empty states. 
 
 Note that Eq.(\ref{hh}) can be rewritten as 
 \beq \label{080}
 | \psi \rangle =\lim_{\theta \to 0} C(|r \rangle -|r+i r \theta \rangle) = \lim_{\theta \to 0} C(|r \rangle -|r+r \theta e^{i \pi/2} \rangle, 
 \eeq  
where we expanded $e^{i \theta}$ around zero. This state reaches the empty state from a fixed phase which is $\pi/2$. To generalize this result, we define this construction  
    \beq  \label{impor}
|E_r \rangle= \lim_{|\Delta r| \to 0} C_r(|r +\Delta r \rangle -|r \rangle)= \lim_{|\Delta r| \to 0} C_r(|r +|\Delta r|e^{i \Delta \theta} \rangle -|r \rangle). 
\eeq 
 This construction is built on adding to $r$ an arbitrary complex variable $\Delta r$. Also, this state always gives a new state since it is just a generalization of Eq.(\ref{080}).  Therefore, this construction is the general formulation of the  empty state, $|E_r \rangle$, and it can be used as another definition of the empty state. The label $E$ refers to the empty state, $r$ is its variable, and $\Delta r$ is a complex variable which we name the source of the empty state.  This expression approaches the empty state from an arbitrary phase, since the source $\Delta r$ can carry any phase. The formulation of Eq.(\ref{impor}) is named as {\em the source representation} of the empty state.

 We note  that any other construction of the empty state rather than Eq.(\ref{impor}) becomes either equivalent or a special case of it. For example, we may reach the empty state from this state
\[
 |\psi_1 \rangle = \lim_{|\Delta r| \to 0} C_1( |r-\Delta r \rangle -| r\rangle ),  
\]
or 
\[
 |\psi_2 \rangle = \lim_{|\Delta r| \to 0} C_2( |r+\Delta r \rangle -| r -\Delta r \rangle ),  
 \]
or may via other possible constructions. All these possibilities will be finally either an equivalent definition to our definition in Eq.(\ref{impor}) or a special case of our definition, as mentioned above.

  Now let us study the empty state from Eq.(\ref{impor}). The normalization constant $C_r$ is given as 
\beq  \label{e4}
C_r= \lim_{|\Delta r |\to 0} \frac{ 1}{\sqrt{2- \langle  r+\Delta r | r \rangle- \langle r | r+\Delta r \rangle}}.
\eeq
 Expanding the inner product  $\langle r | r +\Delta r \rangle$ around  $|\Delta r| \approx 0$. We find  
\beq 
\langle r | r +\Delta r \rangle \approx 1+ g_1(r) \, |\Delta r|-\frac{g_2(r)}{2} |\Delta r|^2+ \cdots,
\eeq
and its complex conjugate 
\beq 
\langle r +\Delta r| r \rangle \approx 1+ g_1^*(r) \, |\Delta r|-\frac{g_2^*(r)}{2} |\Delta r|^2- \cdots, 
\eeq
and their combination is 
\beq \label{ua}
\langle r+\Delta r | r \rangle +c.c. \approx 2+2  \, \textnormal{Re}(g_1(r)) | \Delta r|- \textnormal{Re}(g_2(r)) |\Delta r|^2+ \cdots
\eeq
If $g_1$ is a purely imaginary expression, Eq.(\ref{ua}) becomes 
\beq \label{ewe}
\langle r+\Delta r | r \rangle +c.c. \approx 2- \textnormal{Re}(g_2(r)) |\Delta r|^2+ \cdots
\eeq
Then, substituting Eq.(\ref{ua}) into the normalization constant $C_r$ gives 
\beq \label{xxx}
C_r=\lim_{|\Delta r |\to 0} \frac{ 1}{\sqrt{2- 2  \, \textnormal{Re}(g_1(r)) | \Delta r|- \textnormal{Re}(g_2(r)) |\Delta r|^2}}.
\eeq 
 Now, because the expansion of this expression $|r +\Delta r \rangle - | r \rangle$ in Eq.(\ref{impor}) around $|\Delta r| \approx 0$ is in order of $| \Delta r|$, the expression of $\textnormal{Re}(g_1(r))$ in Eq.(\ref{xxx}) has to be zero to obtain a nonzero empty state. In other words, $\textnormal{Re}(g_1(r))=0$ is the necessary condition for constructing a nonvanishing empty state. Applying this condition, the normalization constant $C_r$ becomes 
\beq \label{eq}
C_r=\lim_{|\Delta r| \to 0} \sqrt{ \frac{ 1}{\textnormal{Re}(g_2(r))}} \, \,  \frac{1}{|\Delta r|}.
\eeq
Substituting this expression back into the empty state $|E_r \rangle$ yields 
\beq  \label{em2}
| E_r \rangle = \sqrt{ \frac{ 1}{\textnormal{Re}(g_2(r))}} \lim_{|\Delta r| \to 0}\frac{ |r+\Delta r \rangle -| r \rangle }{|\Delta r|}.
\eeq
Expanding the state $| r+\Delta r \rangle$ around $| \Delta r| \approx 0$ gives 
\beq \label{ttt}
|r+\Delta r \rangle \approx |r \rangle+ \left( \frac{ \p | r+\Delta r \rangle}{\p |\Delta r|} \right) \bigg|_{|\Delta r|=0}|\Delta r|+\cdots. 
\eeq
Substituting Eq.(\ref{ttt}) into the empty state Eq.(\ref{em2}) yields 
\beq  \label{em}
|E_r \rangle =\sqrt{ \frac{1}{\textnormal{Re}(g_2(r))}} \left( \frac{ \p | r+\Delta r \rangle}{\p |\Delta r|} \right) \bigg|_{|\Delta r|=0}. 
\eeq  
The expression of Eq.(\ref{em}) is the final stage to determine the empty state described by one variable. This formula is equivalent to Eq.(\ref{impor}), and it provides another way to calculate the empty states.  Note that Eq.(\ref{em}) requires two inputs, the state $| r+\Delta r\rangle$ and $g_2(r)$ which is come from expansion of this inner product $\langle r| r+\Delta r \rangle$.

  The formula of empty state in Eq.(\ref{em}) cannot be simplified any further unless we assume new constrains. Such a constrain could be to assume that both $r$ and $\Delta r$ are real, so we can simplify Eq.(\ref{em}) or Eq.(\ref{em2}) as
\beq  \label{ep}
| E_r \rangle = \sqrt{ \frac{ 1}{\textnormal{Re}(g_2(r))}} \left( \frac{d | r \rangle}{d |r|} \right)
\eeq
This state just emphasizes our previous statement  that the empty state $|E_r \rangle$ always gives a new state, which differs from its original state $| r \rangle$. 

 From Eq.(\ref{em}) or Eq.(\ref{ep}), we can see that the empty state may be understood as an operator acting on the original state $| r \rangle$. Therefore, we can generally write any empty state as 
\beq \label{u33}
|E_r \rangle= \hat{A}_r\, |r \rangle, 
\eeq
where $\hat{A}_r$ is the generator operator of the empty state from the original state $| r \rangle$. Therefore, the empty state may be understood as an operator that depends on the variable $r$.  The formulation of  Eq.(\ref{u33}) is considered  {\em the operator representation} of the empty states.

Lastly, we may introduce a third representation of the empty state. Before doing so, we recall the definition of the empty state Eq.(\ref{impor}) 
\[
|E_r \rangle= \lim_{|\Delta r| \to 0} C_r(|r +\Delta r \rangle -|r \rangle)
\]
When $|\Delta r|$ goes to zero, both states will be identical. One can ask why this state does not give a zero? The reason is that the normalization constant $C_r$ as found in Eq.(\ref{eq}) is balanced off the zero of subtraction of the two identical states. Therefore, we should be able to write the empty state as 
 \beq \label{VV}
|E_r \rangle=  \aleph (|r\rangle -|r \rangle ). 
\eeq  
Because this state explicitly represents the state $| r \rangle$, and  it cannot be written as the formulas of either Eq.(\ref{em}) or Eq.(\ref{ep}),  the coefficient $\aleph$ has to be undetermined. We name the formulation of Eq.(\ref{VV}) {\em the self-superposition representation} of empty state.

\subsection{Many Variables}

The above discussion of the empty state is for a single variable case. Here we are going to generalize it to many variables. Suppose that we have a state with $N$-variables described as   $| r_1, r_2, \cdots, r_i, \cdots, r_N \rangle$.  Then, The empty state of the variable $r_i$ can be written as 
\beq \label{i2i}
|E_{r_i} \rangle = \lim_{|\Delta r_i| \to 0} {\bf{C_{r_i}}} (|r_1, r_2, \cdots, r_i+\Delta r_i, \cdots, r_N \rangle -|r_1, r_2, \cdots, r_i, \cdots, r_N \rangle), 
\eeq
where ${\bf C_{r_i}}$ is the normalization constant, function of all variables.  Following a similar procedure as in  our previous discussion of the empty state of one variable, we can write this empty state as  
\beq \label{14}
|E_{r_i} \rangle = \sqrt{\frac{1}{\textnormal{Re}({\bf g_{r_i}})}} \lim_{|\Delta r_i| \to 0} \frac{ \p |r_1,r_2, \cdots, r_i+\Delta r_i, \cdots,r_n \rangle }{\p |\Delta r_i|}, 
\eeq
where ${\bf g_{r_i}}$ came from this inner product expansion
\beq   \label{15}
\langle r_1, r_2, \cdots, r_i+\Delta r_i, \cdots, r_N | r_1,r_2, \cdots, r_i, \cdots, r_N \rangle +c.c. \approx 2 -{\textnormal Re}({\bf g_{r_i}}) |\Delta r_i|^2+ \cdots.
\eeq
 The expression in Eq.(\ref{14}) is the general case of the empty states when the necessary condition is satisfied. To conclude this section, the self-superposition state is named the empty state, and it can be represented by three different representations, namely,  the source, the operator, and the self-superposition representations.  In the next section,  we give an elementary example of empty states and introduce the empty state of the Fock states.

%%%%%%%%%%%%%%%%%%%%%%%%%%%%%%%%%%%%%%%%%%%%%%%%%%%%%%%%%%%%%%%%%%%%%%%%%%%%%%%%%%%%%%%%%%%%%%%%%%%%%%%%%%%%%%%%%%%%%%%%%%%%%%%%%%%%%%%%%%%%

\section{An Elementary Example And The Empty State of The Fock States} 
In the previous section, we defined the empty state as Eq.(\ref{impor}). In this section, we provide an elementary example state of the empty state and find its representations. We also introduce the empty state of the Fock state (empty-Fock state).  

\subsection{The Elementary Example}
Let us start with an elementary example by defining  the state $| R \rangle$ as 
\beq
\label{ee1}
|R \rangle=F (|n\rangle+ R \, |m \rangle),
\eeq 
where $R$ is assumed to be real for simplicity, $F$ is the normalization constant, and $|n \rangle$ and $|m \rangle$ are some number states, where $n \neq m$.  The normalization constant $F$ is found to be  
\beq
F=\frac{1}{\sqrt{1+R^2}}.
\eeq
 The probabilities of finding the state $| n \rangle$ or $| m \rangle$ are given as 
 \beq \label{bn}
 P_n=| \langle n |R \rangle|^2=\frac{1}{1+R^2}, \, \, \, \, \, \, \, \textnormal{and} \, \, \, \, \, \, \, \, \, P_m= |\langle m | R \rangle|^2=\frac{R^2}{1+R^2}. 
 \eeq
 Next, the state $| R+ \Delta R \rangle$ can be written as 
\beq \label{ee3}
|R+\Delta R \rangle=\frac{ |n\rangle+ (R+\Delta R) \, |m \rangle}{\sqrt{1+ (R+\Delta R)^2}}. 
\eeq
Note that $R$ is real and $\Delta R$ also is real. Then, we can find the expression of this inner product and its complex conjugate  as
\beq 
\langle R +\Delta R |R \rangle +c.c. \approx 2- \frac{ (\Delta R)^2}{(1+ R^2)^2} + \cdots
\eeq
This expression is similar to Eq.(\ref{ewe}), where $\textnormal{Re}(g_2(r))$ here is $1/(1+R^2)^2$, so we can write the empty state of the state $|R \rangle $ as 
\beq \label{eae}
|E_R \rangle = \lim_{\Delta R \to 0} C_R(|R+\Delta R \rangle - | R \rangle )= (1+R^2) \frac{ d | R \rangle}{d R}. 
\eeq 
 Expanding this state by using both  definitions  $| R \rangle$ and $F$ gives
\beq \label{efr}
|E_R \rangle = \frac{ -R | n \rangle + |m \rangle}{\sqrt{1+R^2}}
\eeq

This state is the empty state of the $| R \rangle$ state. This state is indeed a new state, since it is a different state than $| R \rangle$. To see that  we calculated the probabilities of finding the state $| n \rangle$ or $| m \rangle$ 

\beq   \label{ee7}
P_n = |\langle n|E_R\rangle|^2= \frac{R^2}{1+R^2}, \, \, \, \, \, \, \, \, \textnormal{and} \, \, \, \, \, \, \, \, \, 
P_m=|\langle m|E_R \rangle|^2= \frac{1}{1+R^2}.
\eeq
These two probabilities in Eq.(\ref{ee7}) are indeed different than the probabilities of Eq.(\ref{bn}). In fact, the two probabilities are switched.  To emphasize that the state $| R \rangle$ and its empty state are not the same, we calculate this inner product 
\beq
\langle R | E_R \rangle = \langle E_R | R \rangle =0. 
\eeq
The two states are orthogonal to each other, so they are different.  

Then, the empty state of $| R \rangle$ state can be represented by using the operator representation (see Eq.(\ref{u33})) as  
\beq 
|E_R \rangle = \hat{A}_R |R \rangle =(1+R^2)\frac{d|R \rangle}{d R}=\left[ -R \, |n \rangle \langle n| +\frac{1}{R} | m \rangle \langle m| \right] |R\rangle, 
\eeq
where $\hat{A}_{R}$ is the generator operator of the empty state $|E_R \rangle$, and it can be written explicitly as 
\beq 
\hat{A}_R=(1+R^2)\frac{d}{d R}= \left[ -R \, |n \rangle \langle n| +\frac{1}{R} | m \rangle \langle m| \right]. 
\eeq
From the above expression we can see that this operator has two expressions. Finally, the self-superposition representation of $| E_R \rangle$ state can be written as 
\beq 
|E_R \rangle = \aleph (|R \rangle - | R \rangle ).  
\eeq

\subsection{The Empty State of The Fock State (Empty-Fock State)}   
In this section, we discuss the empty state of the Fock state (Empty-Fock state). The Fock states or the number states are represented as $|n \rangle$ state, where $n$ can be any positive  integer number. Because the number $n$ is an integer, we cannot apply Eq.(\ref{em}) to find an expression to describe the empty state of the Fock states. However, we can solve this problem by using  an auxiliary state. The auxiliary state is a state that has a continuous variable, say $r$, so when taking  this variable to zero or infinite, we are left with a number state. The state $| R \rangle$ can be an auxiliary state, since when $R$ goes to zero, we left with  $| n \rangle$ state, and when $R$ goes to $\infty$, we left with  $| m \rangle$ state (see Eq.(\ref{ee1})).    

Using the state $| R \rangle$ as an auxiliary state, the empty-Fock state can be represented using the self-superposition representation as  
\beq  \label{5e5}
|E_n \rangle=| E_{R=0} \rangle =\lim_{R \to 0} \aleph(|R \rangle-|R \rangle) =\aleph \, (|n \rangle -  | n \rangle ) = | m\rangle,  
\eeq
where we replace the state $| R=0 \rangle $ to its corresponding state $| n \rangle$ from Eq.(\ref{ee1}), and the resultant state of $|E_{R=0} \rangle$ is $| m \rangle$ state. The state in Eq.(\ref{5e5})  tells us that there is strong dependence on the source state, the auxiliary state $| R \rangle$. If we use a different state of $| R \rangle$ having a different value of $m$, and the same value of $n$, we will get another result for the empty-Fock state, because we have two different sources states. Note that $m$ has to be not equal to $n$, as if they are equal we cannot reach this conclusion. Therefore, using the self-superposition representation for empty-Fock states gives us no clear state. Then, using the source representation, however,  we can write Eq.(\ref{5e5}) as 
\beq 
|E_n \rangle = \lim_{R=0} \lim_{|\Delta R| \to 0} C_{R} (|R + \Delta R \rangle- |R \rangle ). 
\eeq
This representation clarifies  the source state $|R \rangle$ and the source  $\Delta R$.

 The last task we need to perform here is when $R$ being generally any complex value, will Eq.(\ref{5e5}) be invariant? Note, we cannot use the derivative to $| R \rangle$ as we did in Eq.(\ref{eae}). We need to use the general definition of the empty state Eq.(\ref{impor}) or Eq.(\ref{em}). From the LHS of Eq.(\ref{eae}), the normalization constant $C_R$ is found to be 
\beq 
 C_R=\frac{1}{\sqrt{\left[ \frac{1}{\sqrt{1+|R+\Delta R|^2}}-\frac{1}{\sqrt{1+|R|^2}} \right]^2+\left|\left[ \frac{R+\Delta R}{\sqrt{1+|R+\Delta R|^2}}-\frac{R}{\sqrt{1+|R|^2}}\right]\right|^2}}.
\eeq
Because we are interested in the Fock states, we need to use two limits $|\Delta R| \to 0$ and $R \to 0$. After we carriy on the calculations of the two limits, we obtain  
\beq 
|E_n \rangle = \lim_{R\to 0} \lim_{|\Delta R| \to 0} C_R (|R+\Delta R \rangle - |R \rangle )= \aleph (|n\rangle-|n \rangle)=  e^{i \Delta \theta} |m \rangle, 
\eeq 
where $\Delta \theta $ is the phase of $\Delta R$, means $\Delta R= |\Delta R|e^{i \Delta \theta}$. We got the same state $| m \rangle$ of the real $R$ treatment in Eq.(\ref{5e5}). However, the additional thing here is the external phase, so the state is undermined and the complex amplitude as well. 

The last thing we want to say in this section is that although we found that the empty-Fock state is a pure state of  a Fock state, we cannot guarantee that this will be always the case. Maybe if we used another auxiliary state to find the empty-Fock state, we could obtain a mixed state, so the general statement  that can be said now is that the empty-Fock state  highly depends on its source state, its auxiliary state. In the next section, we talk about the empty state of the coherent state.

%%%%%%%%%%%%%%%%%%%%%%%%%%%%%%%%%%%%%%%%%%%%%%%%%%%%%%%%%%%%%%%%%%%%%%%%%%%%%%%%%%%%%%%%%%%%%%%%%%%%%%%%%%%%%%%%%%%%%%%%%%%%%%%%%%%%%%%%%%

\section{The Empty State of The Coherent State (EC State)}
%%%%%%%%%intor 
It is known that the coherent state is a quantum state, and it is the closest state to classical physics  \cite{Glauber ,to,91}. This state is used to build some nonclassical states and is applied to an enormous number of applications (for example,  see\cite{JJ, XI,  HP, HPr, WS, VB}). In this section, we employ the empty state to the coherent state to produce what we call {\em the empty state of the coherent state}, or shortly the (EC state). In this section we study this state and find its mathematical and physical properties.  
%%%%%%%%%%%%%%%%% EC state 
\subsection{The EC State}
The coherent state is defined as the solution of this equation $\hat{a} | \alpha \rangle = \alpha |\alpha \rangle $, where $| \alpha \rangle $ is the coherent state, which can be written as 
\beq  \label{EC1}
|\alpha \rangle= e^{-\frac{|\alpha|^2}{2}} \sum_{n=0}^{\infty} \frac{\alpha^n}{\sqrt{n!}} \, |n \rangle
\eeq
Now, the definition of the empty state of the coherent state or the EC state can be found from Eq.(\ref{impor})
\beq 
|E_\alpha \rangle= \lim_{|\Delta \alpha| \to 0} C_\alpha (|\alpha +\Delta \alpha \rangle - | \alpha \rangle ) 
\eeq 
The inner product of two general coherent states $| \alpha \rangle$ and $|\beta \rangle $ is given as
\beq  \label{EC2}
\langle \beta | \alpha \rangle +c.c. = e^{- \frac{ 1}{2}|\alpha|^2-\frac{1}{2}|\beta|^2+(\beta^*\alpha+\beta \alpha^*)}
\eeq 
Using this relation, we find the inner product of  $\langle \alpha+ \Delta \alpha | \alpha \rangle$ and its complex conjugate as 
\beq 
\langle \alpha+ \Delta \alpha | \alpha \rangle+c.c.=2 e^{-\frac{|\Delta \alpha|^2}{2}} \cos(|\Delta \alpha| |\alpha| \sin(\Delta \theta -\theta)). 
\eeq
This expression can be expanded around $|\Delta \alpha| \approx 0$ to give  
\beq 
\langle \alpha+ \Delta \alpha | \alpha \rangle+c.c. \approx 2- \left[ 1+ |\alpha|^2 \sin(\Delta \theta-\theta)^2\right] |\Delta \alpha|^2 +\cdots, 
\eeq
where $\Delta \alpha$ is expressed as $|\Delta \alpha| e^{i \Delta \theta}$ and $\alpha$ is written as $|\alpha| e^{i \theta}$. Since the linear term of $|\Delta \alpha|$ is absence, that means the necessary condition is satisfied, so the EC state has a nonvinshing empty state. This equation is similar to Eq.(\ref{ewe}) with $g_2$ equal to $\left[ 1+ |\alpha|^2 \sin(\Delta \theta-\theta)^2\right]$, so we can write the EC state as 
\beq \label{EC4}
|E_\alpha \rangle = \frac{1}{\sqrt{1+ |\alpha|^2 \sin(\Delta \theta-\theta)^2}} \lim_{|\Delta \alpha| \to 0} \frac{ |\alpha +\Delta \alpha \rangle - |\alpha \rangle }{|\Delta \alpha| }
\eeq
Now, let us substitute the expression of $| \alpha \rangle$ from Eq.(\ref{EC1}) into Eq.(\ref{EC4}). We then find that 
\beq  \label{EC5}
| E_\alpha \rangle = \frac{1}{\sqrt{1+ |\alpha|^2 \sin(\Delta \theta-\theta)^2}} \lim_{|\Delta \alpha| \to 0} \sum_{n=0}^\infty  \frac{1}{\sqrt{n !} |\Delta \alpha|} \left[ e^{-\frac{1}{2}|\alpha+\Delta \alpha|^2}  (\alpha+\Delta \alpha)^n -e^{-\frac{1}{2} |\alpha|^2} \alpha^n   \right] |n \rangle.
\eeq
The expression inside the bracket inside the summation can be further expanded as 
\beq 
e^{-\frac{1}{2} |\alpha|^2} \left[ e^{- \frac{1}{2} |\Delta \alpha|^2- |\alpha| |\Delta \alpha| \cos(\Delta \theta - \theta) }\left( |\alpha| e^{i \theta} + |\Delta \alpha| e^{i \Delta \theta} \right)^n- |\alpha|^n e^{i n \theta} \right]. 
\eeq
This expression can be expanded around $| \Delta \alpha| \approx 0$. It gives 
\beq 
 e^{-\frac{1}{2} |\alpha|^2} \alpha^n \left[ \frac{e^{i(\Delta \theta - \theta)} n }{|\alpha|}- |\alpha|  \cos(\Delta \theta- \theta) \right] |\Delta \alpha| +\cdots= \left( \frac{ \p | \alpha+\Delta \alpha \rangle}{\p |\Delta \alpha|} \right) \bigg|_{|\Delta \alpha|=0}|\Delta \alpha|+\cdots. 
  \eeq
 By substituting this expression back into Eq.(\ref{EC5}),  the source amplitude $| \Delta \alpha|$ will cancel out from the denominator and numerator. We are left with 
 \beq  \label{0m}
 | E_\alpha \rangle =e^{-\frac{1}{2}|\alpha|^2} \sum_{n=0}^\infty\frac{\alpha^n \left[ \frac{e^{i(\Delta \theta - \theta)} n }{|\alpha|}- |\alpha|  \cos(\Delta \theta- \theta) \right] }{\sqrt{ n !} \sqrt{1+ |\alpha|^2 \sin(\Delta \theta-\theta)^2}} | n \rangle. 
 \eeq
 This expression can be further simplified, if we use this derivation 
 \beq  \label{EC7}
 \frac{ \p | \alpha \rangle }{\p | \alpha|}= e^{- \frac{1}{2} |\alpha|^2}  \sum_{n=0}^\infty  \frac{\alpha^n}{\sqrt{n!}} \left[ \frac{ n}{ |\alpha|} - |\alpha| \right] | n \rangle = \frac{e^{- \frac{1}{2} |\alpha|^2} }{|\alpha|} \sum_{n=0}^\infty \frac{ n \alpha^n}{\sqrt{n!}} \, |n \rangle - |\alpha| | \alpha \rangle, 
 \eeq
 which can be rewritten as 
 \beq \label{GGR}
e^{- \frac{1}{2} |\alpha|^2} \sum_{n=0}^\infty \frac{ n \alpha^n}{\sqrt{n!}} \, |n \rangle = |\alpha| \frac{ \p | \alpha \rangle }{\p | \alpha|}+ |\alpha|^2 | \alpha \rangle
  \eeq
 Then, the EC state can be written as 
 \beq \label{EC8}
 | E_\alpha \rangle = \frac{e^{i\delta \theta}\frac{\p | \alpha \rangle }{\p |\alpha|}+ i |\alpha| \sin\left( \delta \theta\right) |\alpha \rangle  }{\sqrt{1+ |\alpha|^2 \sin(\delta \theta)^2}},
 \eeq  
 where $\delta \theta$ is the phase difference and equals $\Delta \theta-\theta$.  This equation is the complete expression of the EC state. As we can see, the EC state is a new state and differs from the coherent state. Also, the effect of the phase difference $\delta \theta$  can change the physical quantities, as we will see. That means the source's phase $\Delta \theta$ is matter in order to reach the EC state . To illustrate this fact, suppose we reach the EC state from $\Delta \alpha$ having the same phase of $\alpha$, means $\theta= \Delta \theta$, or $\delta \theta=0$. The expression in Eq.(\ref{EC8}) simplifies to be  just the first  derivation. Therefore, maybe we can say that the EC state is a degenerate state. 
 
  Note that if we have more than one EC state, we need to write $\delta \theta$ in the label of the EC state  as $| E_{\alpha, \delta \theta} \rangle$. However, in this article, we do not have two different EC states at the same time, so we do not need to write the angle $\delta \theta$ in the label of the EC state.  
 
 We can define a parameter that measures the ratio of the new state to the original state (here the derivation is the new state). We name this parameter {\em the emptiness} $E$. The emptiness is the ratio of the absolute coefficient of the new state (derivation state) to the absolute coefficient of the original state (coherent state): 
 \beq \label{EE}
 E= \frac{|e^{i \delta \theta}|}{|i |\alpha| \sin(\delta \theta) |}= \frac{1}{|\alpha| \sin(\delta \theta)}. 
 \eeq
If this parameter is infinity, that means it is purely a new state, and if it gives $0$, that means it is a purely coherent state. It becomes infinite when $\delta \theta$ becomes $0$, so this case is the maximum emptiness. Also, the emptiness becomes almost zero when $|\alpha| \gg 1$, and that case is the minimum emptiness. For a given value of $|\alpha|$, the minimum emptiness occurs when $\delta \theta$ equals $\pi/2$.

Also, note that when $\alpha$ goes to zero in the coherent state, we get the vacuum state $| 0 \rangle$. However, when $\alpha$ goes to zero in the EC state, we get one photon state $| 1 \rangle$. Thus, by using the self-superposition representation, we can write 
\beq \label{ja}
|E_0 \rangle =\aleph (|0 \rangle -|0 \rangle)= \lim_{\alpha \to 0} \lim_{|\Delta \alpha| \to 0} C_\alpha (|\alpha +\Delta \alpha \rangle- |\alpha \rangle)= |1 \rangle. 
\eeq
This result agrees with our previous discussion about Eq.(\ref{5e5}), which is that the self-superposition of Fock-states gives us a new state. In the next section, we talk about some of the mathematical properties of EC state.

%%%%%%%%%%%%%%%%%%%%%%% average EC state 

  %%%%%%%%%%%%%%%%%%%%%%%%%%%%%%%%%%%

 \subsection{Identities and mathematical properties }  
  
 In this section, we find some of the mathematical properties of the EC state. First, let us find the operator representation of the EC state, which means we want to find the generator operator $\hat{A_\alpha}$ of the EC state. If we add a photon to the coherent state by applying the operator $\hat{a}$, then the resulting state is given as \cite{mn}
 \beq \label{ECX}
 \hat{a}^\dagger |\alpha \rangle =\left(  \frac{ \partial }{\partial \alpha}+ \frac{ \alpha^*}{2} \right) |\alpha \rangle= \frac{e^{-\frac{1}{2} |\alpha|^2}}{\alpha} \sum_{n=0}^{\infty} \frac{\alpha^n n}{\sqrt{n!}} \, | n \rangle. 
 \eeq
Note here the derivation is with respect to $\alpha$ and not to $|\alpha|$ as we had in the EC state. The relationship between the two derivatives is given as 
\beq 
\frac{ \p |\alpha \rangle}{\p \alpha}= \frac{|\alpha|}{\alpha} \frac{ \p |\alpha \rangle}{\p |\alpha|}+ \frac{ |\alpha|^2}{2 \alpha} |\alpha \rangle,
\eeq
where we can see the difference between the two derivatives. Then, by taking  advantage of Eq.(\ref{GGR}), we can simplifiey the derivation in Eq.(\ref{EC7}) as 
\beq \label{EC9}
\frac{ \p |\alpha \rangle }{\p |\alpha|}=\left( \frac{ \alpha}{|\alpha|} \hat{a}^\dagger  - |\alpha| \right) | \alpha \rangle =\left( \frac{ \alpha}{|\alpha|} \hat{a}^\dagger  - \frac{|\alpha|}{\alpha} \hat{a} \right) | \alpha \rangle 
\eeq
 Then, using this equation allows us to write Eq.(\ref{EC8}) as 
 \beq 
  | E_\alpha \rangle = \left[\frac{e^{i \delta \theta} \frac{\alpha}{ |\alpha|} \hat{a}^\dagger-[1-i \sin(\delta \theta)] \frac{|\alpha|}{\alpha} \hat{a} }{\sqrt{1+ |\alpha|^2 \sin(\delta \theta)^2}} \right] |\alpha \rangle = \hat{A_\alpha} |\alpha \rangle. 
 \eeq  
 The expression inside the bracket is the generator operator $\hat{A_\alpha}$. In case $ \delta \theta$  equals $0$, this operator becomes 
 \beq 
 \hat{A_\alpha}=\left( e^{i \theta} \hat{a}^\dagger- e^{-i \theta} \hat{a} \right),
 \eeq
where again $\theta$ is the complex phase of $\alpha$. This operator in this particular case is a skew-hermitian operator, and commutes with itself.  A similar operator to $\hat{A_\alpha}$  is introduced by Yuen \cite{Yuen}. His operator is $\hat{b}= \mu \hat{a}+ \nu \hat{a}^\dagger$, where $|\mu|^2-|\nu|^2=1$. This operator is satisfying this relation $[ \hat{b}, \hat{b}^\dagger]=1$. Our operator is different than this operator because we have $\nu^\prime=e^{i \theta}$ and $\mu^\prime=e^{-i \theta}$, which satisfies $|\mu^\prime|^2-|\nu^\prime|^2=0$ and $[\hat{A_\alpha},\hat{A_\alpha}^\dagger]=0$. So it is similar but not the same. Also, if we calculate the commutation relation of the general $\hat{A_\alpha}$, we got 
\beq 
[\hat{A_\alpha}, \hat{A_\alpha}^\dagger]=\frac{ \sin(\delta \theta)^2 }{1+ |\alpha|^2 \sin(\delta \theta)^2}  [\hat{a}, \hat{a}^\dagger].
\eeq
 We also may connect the EC state to other known states. For example, we can connect the EC state to the Agarwal state or photon-added coherent state \cite{GSA} throughout  the annihilation operator as
 \beq 
 a^\dagger |\alpha \rangle  =\sqrt{1+|\alpha|^2} \,  | \alpha , 1\rangle, 
 \eeq
 where $| \alpha, 1 \rangle $ is the Agarwal state.

 %%%%%%%%%%%%%%%%%%%%%%%%%%%%% a and a\dagger act on EC state 

 Secondly, we can  find  some identities of the EC state Eq.(\ref{EC8}). Let us start by the annihilation operator $\hat{a}$ acting on $| E_\alpha \rangle$. First, the annihilation operator acting on the derivation of Eq.(\ref{EC9}). It gives 
 \beq 
 \begin{array}{rcl}
 \hat{a} \left( \frac{\p |\alpha \rangle}{\p |\alpha|} \right) & = & \frac{ \alpha}{|\alpha|} \hat{a} \hat{a}^\dagger | \alpha \rangle- |\alpha| \hat{a} |\alpha \rangle, \\
 & = & \frac{\alpha}{|\alpha|}(\hat{a}^\dagger \hat{a}+1) |\alpha \rangle -|\alpha| \alpha |\alpha \rangle, \\
 & = & \alpha \left( \frac{\alpha}{|\alpha|} \hat{a}^\dagger-|\alpha| |\alpha \rangle \right)+\frac{\alpha}{|\alpha|} |\alpha \rangle, \\
 & = & \alpha \frac{\p |\alpha \rangle }{\p |\alpha |} + \frac{ \alpha}{|\alpha|} |\alpha \rangle.
 \end{array}
 \eeq 
 Then, using this expression into $\hat{a}$ acting on the EC state yields 
 \beq  \label{0q}
  \hat{a} |E_\alpha \rangle =  \alpha | E_\alpha \rangle +\frac{ e^{i \delta \theta} \alpha }{|\alpha| \sqrt{1+ |\alpha|^2 \sin(\delta \theta)^2}} |\alpha \rangle.
  \eeq 
  It gives the same state plus the coherent state. For the creation operator, we have 
  \beq  \label{1q}
  \langle E_\alpha | \hat{a}^\dagger =   \langle E_\alpha| \alpha^*+ \langle \alpha | \frac{  \alpha^*e^{-i \delta \theta}}{|\alpha| \sqrt{1+ |\alpha|^2 \sin(\delta \theta)^2}}. 
  \eeq  
   If $\delta \theta $ equals zero, this identity Eq.(\ref{0q}) is simplified to 
  \beq  
  \hat{a} |E_\alpha \rangle = \alpha |E_\alpha \rangle + \frac{ \alpha}{|\alpha|} |\alpha \rangle. 
  \eeq
  Also, we can derive 
    \beq \label{0q1}
 \hat{a}^2 |E_\alpha \rangle = \alpha^2 |E_\alpha \rangle+ \frac{2 e^{i \delta \theta} \alpha^2}{|\alpha| \sqrt{1+|\alpha|^2 \sin(\delta \theta)^2}} |\alpha \rangle
 \eeq
 
   Third is the eigen-equation of the EC state.  From Eq.(\ref{0q}) and Eq.(\ref{0q1}), we noticed that if we multiply Eq.(\ref{0q}) by $2$ and divide Eq.(\ref{0q1}) by $\alpha$, we get exactly same second term for both equations , meaning  
\beq \label{y1}
2\hat{a} |E_\alpha \rangle = 2 \alpha |E_\alpha \rangle+ \frac{2e^{i \delta \theta} \alpha K}{|\alpha|} |\alpha \rangle.
\eeq
\beq \label{y2}
\frac{\hat{a}^2}{\alpha} |E_\alpha \rangle = \alpha |E_\alpha \rangle +\frac{2e^{i \delta \theta} \alpha K}{|\alpha|} |\alpha \rangle.
\eeq
where $K$ is $1/\sqrt{1+|\alpha|^2 \sin(\delta \theta)^2}$. Then, by subtracting the two equations, we can observe the eigen-equation of EC state 
\beq 
\hat{\mathcal{A}_\alpha} |E_\alpha \rangle \equiv \left( 2 \hat{a}-\frac{\hat{a}^2}{\alpha} \right) |E_\alpha \rangle = \alpha |E_\alpha \rangle 
\eeq
where $\hat{\mathcal{A}_\alpha}$ is $\left( 2 \hat{a}-\frac{\hat{a}^2}{\alpha} \right)$. The EC state is not the only possible eigenfunction of this operator, the coherent state $| \alpha \rangle$  itself is a possible solution as well. Also, the operator $\hat{\mathcal{A}_\alpha}$ is just a special case of the C.Brif operator \cite{hh}.

  %%%%%%%%%%%%%%%%%%%%%%%%%%%  inner product 

Fourth, let us turn to calculate the inner products. Starting from the coherent with the derivation state $\langle \alpha | \left( \frac{ \p |\beta \rangle }{\p |\beta|} \right)$.  By using Eqs.(\ref{EC2}, \ref{ECX}, \ref{EC9}), we can find this product as 
  \beq \label{odc}
  \begin{array}{rcl} 
  \langle \alpha | \left( \frac{ \p |\beta \rangle }{\p |\beta|} \right) & = &\langle \alpha | \left( \frac{\beta}{|\beta|} \hat{a}^\dagger |\beta \rangle - |\beta| |\beta \rangle \right) \\
   & = & \frac{ \beta}{|\beta|} \langle \alpha |  \hat{a}^\dagger |\beta \rangle - |\beta | \langle \alpha | \beta \rangle \\
   & = & \langle \alpha | \beta \rangle \left[ \frac{ \beta \alpha^*}{|\beta|}- |\beta| \right] \\
& = & e^{-\frac{1}{2} |\alpha|^2- \frac{1}{2} |\beta|^2+\beta \alpha^*} \left[ \frac{ \beta \alpha^*}{|\beta|}- |\beta| \right].
 \end{array}
  \eeq 
 Note that if the derivation is for $\alpha$ instead of $\beta$, this inner product becomes zero, which means $\langle \alpha | (\p |\alpha \rangle/\p|\alpha|)= 0$.  Using the relation in Eq.(\ref{odc})  and Eq.(\ref{EC8}), we can derive the inner product of the EC state as 
 \beq  \label{QH}
 \langle \alpha | E_\beta \rangle =  \frac{e^{-\frac{1}{2} |\alpha|^2- \frac{1}{2} |\beta|^2+\beta \alpha^*} }{\sqrt{1+ |\beta|^2 \sin(\delta \theta)^2}} \left[ e^{i \delta \theta} \left( \frac{ \beta \alpha^*}{|\beta|}-|\beta| \right)+i |\beta| \sin(\delta \theta) \right].
 \eeq
 In case both of them are $\alpha$, we get 
 \beq \label{qw}
 \langle \alpha | E_\alpha \rangle = \frac{ i |\alpha| \sin(\delta \theta)}{ \sqrt{1+ |\alpha|^2 \sin(\delta \theta)^2}}  
 \eeq
 
Next, the inner product of two derivation states can be found as 
 \beq 
 \left( \frac{ \p \langle \alpha |}{\p |\alpha|} \right) \left( \frac{ \p | \beta \rangle }{\p |\beta|} \right)= e^{-\frac{1}{2} |\alpha|^2- \frac{1}{2} |\beta|^2+\beta \alpha^*} \left[ \frac{ \alpha^* \beta}{|\alpha| |\beta|}+ \frac{ \alpha^{*2} \beta^2}{|\alpha ||\beta|}-\frac{|\beta| \alpha^* \beta}{|\alpha|}- \frac{|\alpha| \alpha^* \beta}{|\beta|}+|\alpha||\beta| \right].
  \eeq
  This relation can be used to calculate the inner product of the EC states, if we wish. Also, note that when $\beta$ equals $\alpha$, we get $1$. The next section discusses the statistical properties of the EC state.

 %%%%%%%%%%%%%%%%%%%%%%%%%%%%%

 \subsection{Statistical properties} 
 The statistical properties of a photonic state is an excellent way to know the photon distribution, the average number of photons, photons fluctuations, etc. In this section, we find some of the statistical properties of the EC state. Let us start by  studying  the photon probability distribution, which can be found by multiplying Eq.(\ref{0m}) by $\langle n |$ and then square it. We found 
 \beq  \label{P1}
 P_n=|\langle n | E_\alpha \rangle |^2 = \frac{ e^{-|\alpha|^2} |\alpha|^{2 n} \left[ \frac{n^2}{|\alpha|^2}+ \cos(\delta \theta)^2 (|\alpha|^2-2n) \right]}{n! \left( 1+ |\alpha|^2 \sin( \delta \theta )^2 \right)}. 
 \eeq 
This distribution is illustrated in Fig.(1). In the case when $\alpha$ becomes zero, one photon always will survive, which agrees  with Eq.(\ref{ja}).    
   \begin{figure}[h!]
   \begin{center}
\gr[scale=0.4]{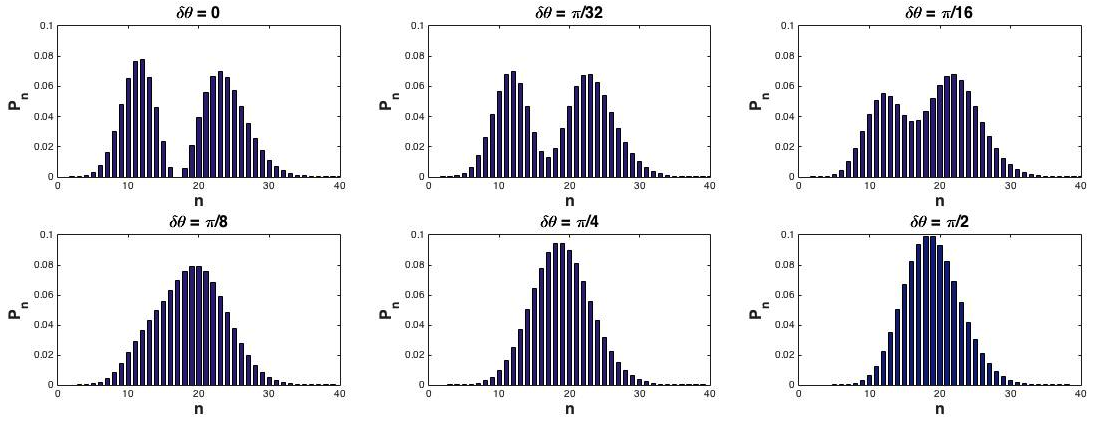} 
\caption{ The photon probability distribution of the EC state for different values of the phase differences $\delta \theta$. The value of $|\alpha|$ here is $4$. }
\end{center}
\end{figure}

 From Fig.(1) we can see one point with a zero probability at the angle $\delta \theta=0$. This zero point occurs for $n=|\alpha|^2$. It indicates the maximum  quantum interference.  This result agrees with the definition of the emptiness in Eq.(\ref{EE}) $E$, where when the emptiness is maximized, the quantum interference is maximized as well. Therefore, the emptiness can be used  to measure the strength of the quantum interference of the EC state.

 %%%%%%%%%%%%%%%%%%%%%%%%%%%%%%%%%%%%%%%% 

 Then,  we calculate the average photon number of the EC state. Using Eqs.(\ref{0q},\ref{1q}), we can calculate the average photon number $\langle \hat{n} \rangle$, which gives 
 \beq  \label{v1}
 \langle \hat{n} \rangle= \langle E_\alpha | \hat{a}^\dagger \hat{a} | E_\alpha \rangle = |\alpha|^2+ \frac{1+2 |\alpha|^2 \sin(\delta \theta)^2}{1+|\alpha|^2 \sin(\delta \theta)^2}= |\alpha|^2+ M, 
  \eeq
 where $M$ is defined as 
 \beq
 M= \frac{1+2 |\alpha|^2 \sin(\delta \theta)^2}{1+|\alpha|^2 \sin(\delta \theta)^2}.
  \eeq
  The maximum value of Eq.(\ref{v1}) is $2+|\alpha|^2$, and that happens when $\delta \theta$ equals $\pi /2$ and $|\alpha|^2 \gg 1$. In contrast,  the minimum value is $|\alpha|^2+1$, which happens when $\delta \theta $ equals $0$. Therefore, this average generally is bounded by an upper and lower bound described as 
 \[ 
 1 \le M \le 2
 \]
 \beq 
 1+|\alpha|^2 \le  \langle \hat{n} \rangle \le 2+|\alpha|^2. 
  \eeq
 This bound indicates that we always have at least one photon on average. 
 
 Then, after using the identity in Eq.(\ref{0q1}), we can find the average of the square photon number operator as
 \beq 
 \langle \hat{n^2} \rangle = \langle \hat{a}^\dagger \hat{a} \hat{a}^\dagger \hat{a} \rangle = |\alpha|^4+ 5 |\alpha|^2+ \frac{1+2 |\alpha|^2 \sin(\delta \theta)^2}{1+|\alpha|^2 \sin(\delta \theta)^2}= |\alpha|^4+5 |\alpha|^2+M.    \eeq
 The photon number fluctuations $\langle \Delta n \rangle$ are given as 
 \beq \label{112}
 \langle \Delta n \rangle= \sqrt{ \langle \hat{n^2} \rangle -\langle \hat{n} \rangle^2}= \sqrt{|\alpha|^2 (5-2 M)+M(1-M)}, 
 \eeq 
 This expression is always real and positive, and it becomes equal to $\sqrt{3} \,  |\alpha|$ when the phase difference $\delta \theta$ equals $0$. And the minimum fluctuations happens when the phase difference equals to $\pi /2$ (see Fig.(2)). 
     \begin{figure}[h!]
   \begin{center}
\gr[scale=0.5]{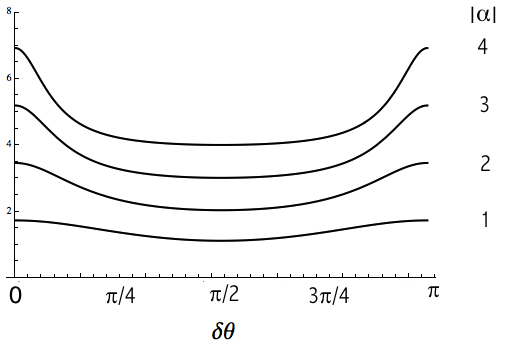} 
\caption{ The photon number fluctuation of the EC state Eq.(\ref{112}) for some values of $|\alpha|$.}
\end{center}
\end{figure}

 %%%%%%%%%%%%%%%%%%%%%%%The mandel 
 
The next statistical property is the classification of the photon number distributions. If we want to know whether the EC state is a sub- or super- Poissonian distribution, we can use the Mandel Q-parameter \cite{yy}. The Mandel Q-parameter is one of the parameters that enables us to know the category of the distribution . It is defined as  
\beq \label{QQ}
Q= \frac{ \langle \Delta \hat{n} \rangle^2- \langle \hat{n} \rangle}{\langle \hat{n} \rangle}.  
\eeq
If $Q$ is $-1 \le Q <0$, the statistics become a sup-Poissonian distribution. Also, if $Q$ is $Q>0$, the statistics become a super-Poissonian distribution, and if $Q$ is $-1$, the statistics become the number state distribution. By substituting these Eqs.(\ref{v1},\ref{112}) into the Mandel Q-parameter Eq.(\ref{QQ}), we find 
\beq \label{u3}
Q=\frac{ 2|\alpha|^2(2-M)-M^2}{|\alpha|^2+M} 
\eeq
This expression exhibits both sub- and super- Poissonian distribution (see Fig.(4)), which means it can be used to produce photon antibunching.

\begin{figure}[h!]
\begin{center}
\gr[scale=0.5]{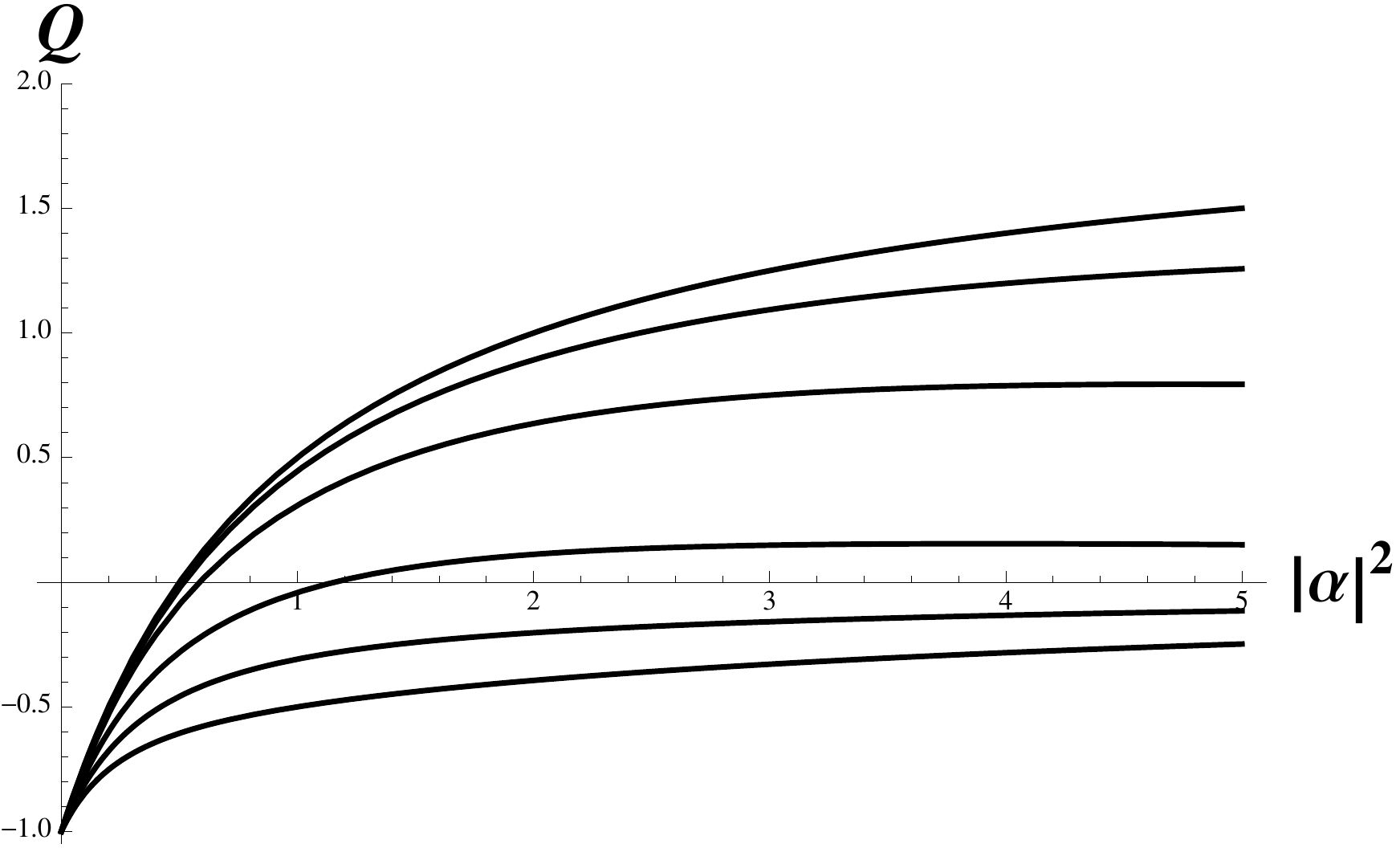} 
\caption{The Mandel Q-parameter of the EC state Eq.(\ref{u3}) for different values of $\delta \theta$. The values of $\delta \theta$  are $0, \pi/20, \pi/10, \pi,5, 3 \pi/10$$,$ and $\pi/2$. The largest curve  for $\delta \theta$ equals $0$, and all other values are arranged, respectively. }
\end{center}
\end{figure}    
 
 %%%%%%%%%%%%%%%%%%%%%%%%%%% phase 
 
 Next, we calculate the quantum phase distribution $\mathcal{P}(\varphi)$ of EC state. We selected the Pegg-Barnett formulation of the phase distribution \cite{111, 222}. This distribution is defined as  
  \beq 
 \mathcal{P}(\varphi)= \frac{1}{2\pi} | \langle \varphi| r \rangle |^2, 
 \eeq
 where the state $|\varphi \rangle$ is defined as 
 \beq 
 | \varphi \rangle = \sum_{n=0}^{\infty} e^{i n \varphi} |n \rangle. 
 \eeq
 Then, using Eq.(\ref{0m}) it is allows us to find this distribution for the EC state
 \beq 
 \mathcal{P}(\varphi)=\frac{1}{2 \pi} |\langle \varphi | E_\alpha \rangle|^2= \frac{e^{-|\alpha|^2}}{2 \pi [1+|\alpha|^2\sin(\delta \theta)^2]} \left| \sum_{n=0}^\infty \frac{e^{-i (n \varphi-\theta)}|\alpha|^n}{\sqrt{n!}}  \left(   \frac{n e^{i \delta \theta}}{|\alpha|} -|\alpha| \cos(\delta \theta)     \right) \right|^2
 \eeq
 This distribution becomes close to the phase distribution of the coherent state when $\delta \theta$ becomes $\pi/2$. In this case, there is one peak. However, when $\delta \theta$ becomes $0$, this distribution gives two peaks. In the next section, we talk about the nonclassicality properties of the EC state.

 %%%%%%%%%%%%%%%%%%%%%

 \subsection{Nonclassical Properites}
 In this section we find some nonclassical properties of the EC state. We started first the squeezing properties and then  the quesi-probability distributions. 
 
 \subsubsection{ Squeezing Properties} 
 %%%%%%%%%%%%%%% the position and momentum operators 
Some of the squeezing properties can be found by finding the quadrature operators.  These quadrature operators $\hat{X_1}$ and $\hat{X_2}$ are the operators of the dimensionless position and momentum. They are defined as 
\beq 
\hat{X_1}= \frac{\hat{a}+ \hat{a}^\dagger}{2}, \, \, \, \, \, \, \, \, \, \, \, \, \, \hat{X_2}= \frac{\hat{a}-\hat{a}^\dagger}{2 i}. 
\eeq
Whenever the fluctuations of one of these two operators of a given state are less than $1/4$, that state is  squeezed. Namely, the condition reads as 
\beq \label{plt}
\langle (\Delta \hat{X_i}) ^2\rangle < \frac{1}{4}. 
\eeq

  Now let us  find the expectation values of these operators for the EC state. Before doing so, let us write certain helpful mathematical relations by using these  Eqs.(\ref{0q},\ref{0q1},\ref{qw}). The helpful relations are
\[ 
\langle E_\alpha | \hat{a} | E_\alpha \rangle = \alpha \left(1-i  K^2 \sin(\delta \theta) e^{i \delta \theta} \right), 
\] 
\[
\langle E_\alpha | \hat{a}^2 | E_\alpha \rangle =\alpha^2\left(1-2i K^2 \sin(\delta \theta) e^{i \delta \theta}\right), 
\] 
\beq 
\langle E_\alpha |\hat{a} \hat{a}^\dagger | E_\alpha \rangle = 1+ |\alpha|^2+ K^2(1+ 2 |\alpha|^2 \sin(\delta \theta)^2 ),
\eeq
where $K$ is the expression $1/\sqrt{1+|\alpha|^2 \sin(\delta \theta)^2}$. Note that $K$ is always bounded between $0$ and $1$. Then, the expectation values of the operators $\hat{X_1}$ and $\hat{X_2}$ for the EC state are given by
\beq 
\langle E_\alpha | \hat{X_1} | E_\alpha\rangle =\frac{ \alpha+ \alpha^*}{2} + |\alpha| K^2 \sin(\delta \theta) \sin( \Delta \theta), 
\eeq
\beq 
\langle E_\alpha | \hat{X_2} | E_\alpha \rangle = \frac{\alpha -\alpha^*}{2i} - |\alpha| K^2 \sin(\delta \theta)\cos(\Delta \theta). 
\eeq
Recall that $\delta \theta$ is defined as $\Delta \theta - \theta$. Therefore, in addition to the phase difference $\delta \theta$,  we need to know the source phase $\Delta \theta$.  We note that if $\Delta \theta$ equals $\theta$, the above expressions tend to give exactly the same results as the coherent states, where the second term becomes zero. 

Next, the expectation values for the square of $\hat{X_1}$ and $\hat{X_2}$ operators are given as
\beq 
\langle E_\alpha | \hat{X_1}^2 | E_\alpha \rangle =\frac{1}{4}+\frac{\alpha^2+\alpha^{*2}}{4}+\frac{|\alpha|^2}{2}+ \frac{K^2}{2} \left[ 1+2 |\alpha|^2 S_1    \right],
\eeq 
where $S_1$ is $\sin(\delta \theta) [ \sin(\delta \theta)+ \sin(2\Delta \theta-\delta\theta)]$ and 
\beq 
\langle E_\alpha | \hat{X_2}^2 | E_\alpha \rangle =\frac{1}{4}-\frac{\alpha^2+\alpha^{*2}}{4}+\frac{|\alpha|^2}{2}+ \frac{K^2}{2} \left[ 1+2 |\alpha|^2 S_2    \right],
\eeq
 where $S_2$ is  $\sin(\delta \theta) [ \sin(\delta \theta)- \sin(2\Delta \theta-\delta \theta)]$. The fluctuations of these operators are given as 
 \beq \label{X1}
 \langle (\Delta \hat{X_1} )^2\rangle =\langle \hat{X_1}^2\rangle-\langle \hat{X_1} \rangle^2= \frac{1}{4}+K^2 \left[ \frac{1}{2}- |\alpha|^2 K^2 \sin(\delta \theta)^2 \sin(\Delta \theta)^2 \right]. 
 \eeq
  \beq \label{X2}
 \langle( \Delta \hat{X_2})^2 \rangle =\langle \hat{X_2}^2\rangle-\langle \hat{X_2} \rangle^2=  \frac{1}{4}+K^2 \left[ \frac{1}{2}- |\alpha|^2 K^2 \sin(\delta \theta)^2 \cos(\Delta \theta)^2 \right]. 
   \eeq  
     \begin{figure}[h!]
   \begin{center}
\gr[scale=0.5]{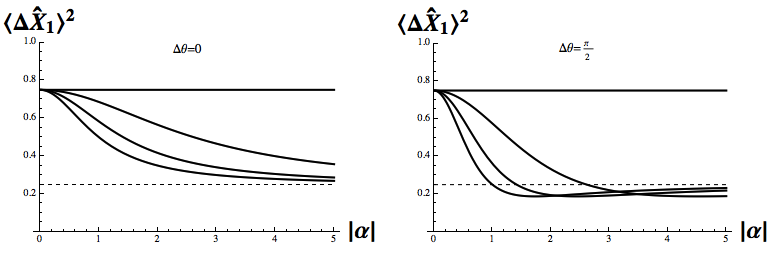} 
\caption{ The fluctuation of $\hat{X_1}$ operator of the EC state for some values of $\Delta \theta$ and $\delta \theta$. Each plot has $4$ values of $\delta \theta$ which are form the top to down $0, \pi/8,\pi/4,\pi/2$. The straight line in all figures is for $\delta \theta=0$.    }
\end{center}
\end{figure}   
In the above expressions, when $\delta \theta$ equals zero, they gives $3/4$. This is the minimum fluctuations of one photon. In Fig.(4) we representes the fluctuations of  $\hat{X_1}$ operator for some angles of $\delta \theta$ and $\Delta \theta$. We can see that the squeezing can occur for some values of $\Delta \theta$. Applying Eq.(\ref{plt})  to Eq.(\ref{X1}) gives us the squeezing condition 
  \beq 
  \frac{1}{4} < |\alpha| K \sin(\delta \theta) \sin(\Delta \theta). 
  \eeq
The maximum squeezing occurs when $\delta \theta$ equals $\pi/2$ and $\Delta \theta$ equals $\pi/2$ or $0$ for $\hat{X_2}$ operaator, and $|\alpha|$ equals $\sqrt{3}$. When that happens, Eq.(\ref{X1}) gives $3/16$, and Eq.(\ref{X2}) gives $3/8$. 

The multiplication of Eq.(\ref{X1}) and Eq.(\ref{X2}) always satisfies these upper and lower bounds  
\beq 
\frac{1}{16} \le \langle \Delta \hat{X_1} \rangle ^2 \langle \Delta \hat{X_2} \rangle^2 \le \frac{9}{16}
\eeq
For example, in case both $\delta \theta$ and $\Delta \theta$ equal $\pi/2$, the multiplication yields 
\beq 
 \langle \Delta \hat{X_1} \rangle ^2 \langle \Delta \hat{X_2} \rangle^2 = \frac{1}{16} \frac{(3+|\alpha|^2)(3+|\alpha|^4)}{ (1+|\alpha|^2)^3}. 
 \eeq
 This relation starts from $9/16$ when $|\alpha|$ is zero, and finishes at $1/16$ when $|\alpha| \gg 1$. Next, we study the qusi-probability distributions for the EC state.

  \subsubsection{ Qusi-Probability Distributions}
 In quantum optics, there are some qusi-probablity distributions that give an indication of whether a light state represents some nonclassical features or does not. The first distribution  is the $Q(\alpha)$, or Husimi function  \cite{tt}. This function gives us an idea of the behaviour of the interested state in the phase space. It is defined as 
 \beq 
 Q=\frac{\langle \alpha |\hat{\rho}| \alpha \rangle }{\pi}, 
 \eeq
 where $\hat{\rho}$ is the density operator. The density operator of the EC state is $| E_\alpha \rangle \langle E_\alpha|$. Applying this density operator to  $Q(\alpha)$ yields $Q(\alpha)=|\langle \alpha | E_\beta \rangle|^2/\pi$. Using the identity in Eq.(\ref{QH}) allows us to calculate the $Q(\alpha)$ distribution  
  \beq \label{qes}
 Q=\frac{K^2}{\pi}e^{-|\alpha-\beta|^2} \left[ |\alpha-\beta|^2 \cos(\delta \theta)^2+|\alpha|^2\sin(\delta \theta)^2+ \frac{\alpha^* \beta-\beta^* \alpha}{2 i} \sin(2 \delta \theta) \right] ,
 \eeq
 where $K$ is $1/\sqrt{1+|\beta|^2 \sin(\delta \theta)^2}$. This function is always positive. In Fig.(5) we plot Eq.(\ref{qes}) for some values of $\delta \theta$.  We plotted $Q(\alpha)$ in Fig.(5) versus the real and imaginary parts of $\alpha=\alpha_r+i \alpha_i$. From the same figure, we can see how $Q(\alpha)$ gives a semi-crescent shape which indicates there is a quantum interference.  
 \begin{figure}[h!]
\begin{center}
\gr[scale=0.4]{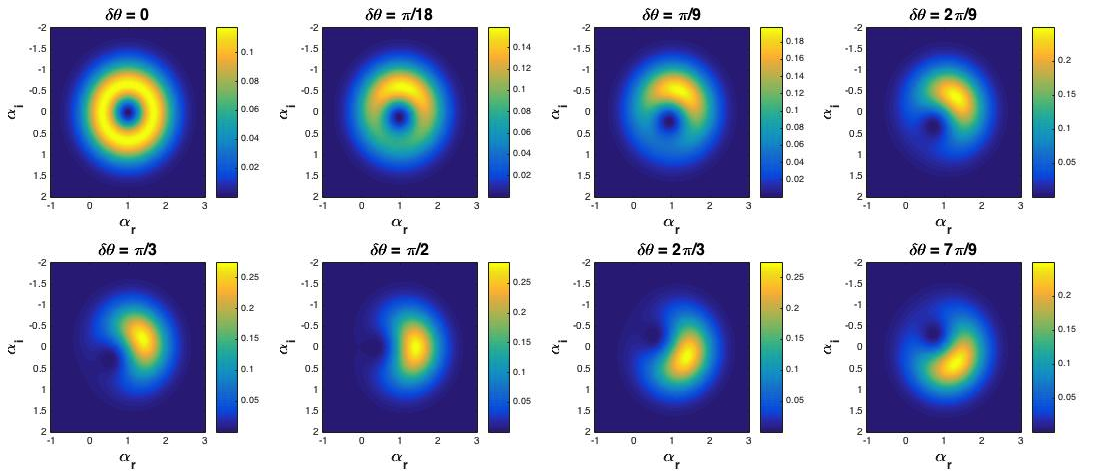} 
\caption{The $Q(\alpha)$ function of the EC state for different phase differences $\delta \theta$. The axis are the real and imaginary parts of $\alpha$. The value of $\beta$ is real and equals $1$. }
\end{center}
\end{figure}   
This nonclassical behaviour is maximized when $\delta \theta$ equals $0$. In this case, the Hsuimi function looks like a ring, and its equation can be reduced to the following 
\beq 
Q(\alpha)=\frac{|\alpha-\beta|^2}{\pi} e^{-|\alpha-\beta|^2}.
\eeq
 Also, we note that as the angle $\delta \theta$ increases, the ring deforms . That deformation is maximized at the angle $\pi/2$.  

The quantum antinormally ordered characteristic function is defined as  
\beq
 C_A(\lambda)= \int d^2\alpha\,  Q(\alpha)e^{\lambda \alpha^*-\lambda^* \alpha}. 
 \eeq
Substituting the expression of $Q(\alpha)$ into this formula yields $C_A(\lambda)$ to be
\[
 C_A(\lambda)=\frac{K^2}{2} e^{\beta^* \lambda-\beta \lambda^*-|\lambda|^2} \bigg[ 2+|\beta|^2+\beta^* \lambda-\beta \lambda^*-2|\lambda|^2-(\beta^* \lambda-\beta \lambda^*+|\beta|^2) \cos(2\delta \theta) +
\]
 \beq
 i (\beta^* \lambda+\beta \lambda^*) \sin(2 \delta \theta) \bigg].  
 \eeq
Using this equation allows us to find the Wigner characteristic function, where it is defined as
 \beq 
 C_W(\lambda)= C_A(\lambda)e^{\frac{1}{2} |\lambda|^2}. 
 \eeq
 This characteristic function enbles us to find the Wigner function $W(\alpha)$. The Wigner function is a quesi-probablity distribution and gives a clear sign for whether a state  has  nonclassical properties or does not. If there are some negative regions in the Wigner function, then the state is a nonclassical state. 
 
 The Wigner function can be defined as 
 \beq
 W(\alpha)= \frac{1}{\pi^2} \int d^2 \lambda C_W(\lambda)e^{\lambda^* \alpha-\lambda \alpha^*}. 
 \eeq
  Using this definition, we find the Wigner function as 
 \[
 W(\alpha)=-\frac{K^2}{\pi}e^{-2|\alpha-\beta|^2}\bigg[ 2-8 |\alpha|^2-|\beta|^2(5+3 \cos(2 \delta \theta))+(\alpha^* \beta+\alpha \beta^*)(6+2 \cos(2 \delta \theta))+
\] 
\beq \label{yq}
2i (\alpha^* \beta-\alpha \beta^*) \sin(2 \delta \theta)\bigg]. 
\eeq
 \begin{figure}[h!]
\begin{center}
\gr[scale=0.4]{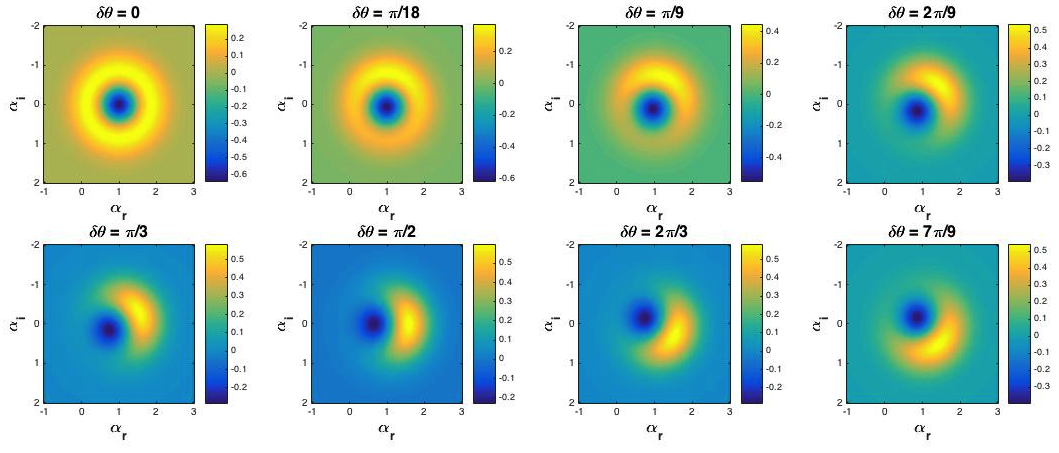} 
\caption{The Wigner  function $W(\alpha)$ of the EC state for different phase differences $\delta \theta$. The value of $\beta$ is real and quals $1$. }
\end{center}
\end{figure}   
When $\delta \theta$ equals zero, this function reduces to 
\beq 
W(\alpha)= \frac{2}{\pi} (4|\alpha-\beta|^2-1)e^{-2|\alpha-\beta|^2}. 
\eeq
The Wigner function is illustrated in Fig.(6). From Eq.(\ref{yq}) and Fig.(6), we can say that the EC state  is always a  nonclassical state. In other words, the Wigner function always has some negative regions. Also, the maximum negative value of $W(\alpha)$ occurs when $\delta \theta$ equals $0$. 

 Final remarks about the nonclassical properties of the EC state. We found that the extreme nonclassical behaviour or higher negativity of the Wigner function is occurs when $\delta \theta$ equals $0$. Also, the squeezing properties and $Q(\alpha)$ function  maximize  the nonclassical behaviour of the EC state when $\delta \theta$ equals $0$. That angle is that same angle which maximize the emptiness in Eq.(\ref{EE}). This means the emptiness is a good way to represent the nonclassical behaviour of EC state. Also, it means the derivation state in Eq.(\ref{EC7}) is by itself a nonclassical state.

%%%%%%%%%%%%%%%%%%%%%%%%%%%%%%%%%%%%%%%%%%%%%%%%%%%%

\section{Conclusion}
The empty states are a new class of states, and we believe that these states are promising. We think that because of their unexpected behaviour starting from the initial thought of them to give nothing to their nonclassical behaviour as for the EC state.  These unexpected properties of the empty states let us believe that these states will find their way in the applications of physics.

%%%%%%%%%%%%%%%%%%%%%%%%%%%%%%%%%%%%%%%%%%%%%%%%

\section{Acknowledgements}
We acknowledge Omar Alshehri  for his interesting remarks, and scientific review. Also, we acknowledge Professor David Yevick for his interesting comments. We further acknowledge financial support from the Taibah University.

ÿ%%%%%%%%%%%%%%%%%%%%%%%%%%%%%%%%%%%%%%%%%%%%%%%%%%%

\end{document}